# Multiple Sequence Alignment is not a Solved Problem

David A. Morrison
Department of Organismal Biology, Uppsala University, Sweden

**Abstract**

Multiple sequence alignment is a basic procedure in molecular biology, and it is often treated as being essentially a solved computational problem. However, this is not so, and here I review the evidence for this claim, and outline the requirements for a solution. The goal of alignment is often stated to be to juxtapose nucleotides (or their derivatives, such as amino acids) that have been inherited from a common ancestral nucleotide (although other goals are also possible). Unfortunately, this is not an operational definition, because homology (in this sense) refers to unique and unobservable historical events, and so there can be no objective mathematical function to optimize. Consequently, almost all algorithms developed for multiple sequence alignment are based on optimizing some sort of compositional similarity (similarity = homology + analogy). As a result, many, if not most, practitioners either manually modify computer-produced alignments or they perform de novo manual alignment, especially in the field of phylogenetics. So, if homology is the goal, then multiple sequence alignment is not yet a solved computational problem. Several criteria have been developed by biologists to help them identify potential homologies (compositional, ontogenetic, topographical and functional similarity, plus conjunction and congruence), and these criteria can be applied to molecular data, in principle. Current computer programs do implement one (or occasionally two) of these criteria, but no program implements them all. What is needed is a program that evaluates all of the evidence for the sequence homologies, optimizes their combination, and thus produces the best hypotheses of homology. This is basically an inference problem not an optimization problem.

## Introduction

Bioinformatics is the nexus of the biological and computational sciences. Therefore, all bioinformatic analyses should solve both a biological and a computational problem. However, most published discussions of these problems occur in the bioinformatics literature, which tends to treat them as almost solely computational issues. Algorithms are important in science because they can provide objective and repeatable procedures for turning observations into inferences; but we first need to understand the scientific goal. The purpose of this paper is specifically to inject some detailed biology into a particular aspect of bioinformatics.

The topic is the multiple sequence alignment problem, which is one of the oldest problems in computational biology, and one of supreme practical importance[1,2]. It is usually claimed to be conceptually important, as well, being related to the biological concept of homology. Yet, in practice, it is often treated merely as a tool, being little more than a 'bottleneck' in sequence-processing pipelines, for which there are scores of computer programs available to carry out the analyses. However, while alignment and homology are equated in principle, in practice this important connection has actually been irrelevant for almost all of the automated sequence alignment programs.

This issue is particularly acute in the field of phylogenetics. Sequence alignments can be used for many different purposes in biology, and not all of these purposes will necessarily be best served by the same alignment. For example, Morrison[3] illustrates 15 different alignments of the same sequences, only one of which is intended explicitly for phylogenetics. Furthermore that paper[3] recognizes that multiple sequence alignment has at least four major strands: sequence comparison (including assessing data quality[4]); database searching; structure prediction; and phylogenetic analysis. While this fragmentation of purpose *has* been recognized in the bioinformatics literature (e.g. Friedrich et al.[5]), the fact that these purposes may require distinct alignments is rarely recognized.

My specific topic in this paper is that the production of multiple sequences alignments explicitly for phylogenetic purposes has not yet been solved computationally. Much progress has been made towards the other three purposes, but phylogenetics lags far behind. In particular, if an alignment truly does represent hypotheses of homology among the characteristics of the organisms, then the alignment itself is also a worthwhile goal, on its own — the alignment can be used for many other purposes than building a phylogenetic tree.

This can be clearly seen in the fact that biologists are often not happy with the output of the current crop of alignment programs. For example, Morrison[6] surveyed 1,280 empirical papers containing phylogenetic analyses, published in 26 biology journals, and noted that 26% of the papers constructed their multiple sequence alignment manually, a further 19% manually modified a computer-based alignment, and a further 8% used some automated procedure to modify the alignment. That is, more than 50% of the papers were unwilling to accept the output of a simple computer-produced alignment procedure as useful for their purposes (i.e. the alignment is a likely representation of homology relationships). That situation has not changed much since then, except in the field of genome alignments, which cannot be constructed manually.

This suggests that the current computer programs (e.g. see the lists of programs[7-9]) do not yet solve the biological part of the multiple sequence alignment problem. They provide many innovative solutions to particular computational problems, but the sum of these solutions still does not solve the overall biological problem. Thus, for the biological goal there is still no single program that can be relied upon.

The basic practical and theoretical issue is that there can be no objective mathematical function for homology, because homology (in the phylogenetic sense) refers to unique and unobservable historical events. Homology exists independently of our ability to recognize it, and therefore its recognition is an inference problem rather than an optimization problem. There is nothing quantitative to optimize, and thus no optimization algorithm for a mathematician to develop. And yet, the computational goal is to produce alignments that no biologist would see any reason to modify for phylogenetic purposes.

In this paper, I review those features of the biological problem that seem to be the most important for computationalists to understand. This will explain why the current computations do not yet adequately address the problem, and identify the biological problem in terms of the required computational goal(s). My argument is that in bioinformatics the biology should come first (in time), the mathematics second, and then the computing. This has not happened for multiple alignment of nucleotide sequences, if the alignments are intended to represent hypotheses of homology. Algorithmic developments have not been explicitly directed towards the known biological criteria for homology, and so the mathematical optimality criteria have not been motivated by the processes responsible for generating the sequence data.

So, I do not attempt to solve the computational problems, although I do outline some feasible approaches. My ultimate aim is to convince computationalists that the multiple sequence alignment problem is still open, and that it should provide some interesting computational challenges. My hope is, therefore, that this paper will motivate more research in this area, as well as providing the necessary biological background for computational people new to it.

The paper is arranged as follows. First, I try to clarify the very specific meaning of homology within phylogenetics, and provide definitions and terms required for the rest of the paper. Next, I provide an introduction to the complex nature of the concept of homology, and the possible evidence for detecting it, thus providing the conceptual background to what the biologists are trying to achieve. Then, I summarize those algorithmic developments that might be relevant to assessing evidence for homology. Finally, I discuss possible approaches to developing a computer program that might provide a practical method of producing a homology alignment

**Homology and Phylogenetics**

In this section, I try to clarify a set of biological terms and their concepts that seem to be rather confused in the literature. I refer to literature sources that contain more detailed discussions of various of these concepts.

*Biology, Linguistics and Stemmatology*

Phylogenetic methods have been developed and are used in at least three quite distinct areas of knowledge: biology (the study of life), historical linguistics (the study of languages), and stemmatology (the study of manuscripts). The history of thought in all three fields is briefly discussed by Morrison[10].

In this paper, I will present the ideas solely in terms of biology. This might seriously restrict the applicability of those ideas, because there are some quite fundamental differences between the three fields, which do have practical consequences. In particular, there is no equivalent of the concept of “genotype” in either linguistics or stemmatology, whereas all three fields do have a concept of “phenotype” (see below).

*Objects and Characters*

For our purposes here, the term **phylogenetics** refers to the study of the genealogy of a set of identifiable **objects** that have observable characteristics. The characteristics, or **characters**, can have any one of several (possibly mutually exclusive) variants, called “states”. Each object will display one (or possibly more) of the states for each character.

The objects are assumed to have a history of descent via some sort of copying. So, there is a common ancestral object with a unique set of character states. Through time, this ancestor gives rise to a set of descendants, which give rise in turn to yet more descendants. As copying proceeds, the characteristics may be modified, so that characters can have an **ancestral state** and one or more **derived states**. Different subsets of the descendant objects will have different combinations of characters with ancestral and derived states.

The objective of phylogenetics is to reconstruct the historical relationships of *both* the objects and the characters.

*Characters*

Much of the confusion about characters occurs because of a failure to clearly distinguish “phenotype” from “genotype”. **Genotype** refers to those characters that are physically inherited from parent to offspring. That is, they are the characters that are actually copied when each descendant is created. **Phenotype** refers to characters that are the result of interactions among the genotype characters, or between the genotype characters and their environment. That is, genotypes are inherited whereas phenotypes are expressed.

We can attempt to establish the historical relationships among the phenotype characters just as much as among the genotype characters. However, it is expected, a priori, that genotype characters will be more reliable for establishing the correct historical relationships among the objects.

Strictly speaking, only the chromosomes are part of the genotype. For our purposes, these consist of DNA, with each chromosome being made up of an ordered string of nucleotides, forming genes. (Aside: in some parts of biology, it is RNA not DNA that is copied.) So, genotype consists of nucleotides and genes. All other characters are, strictly speaking, part of the phenotype[11].

This often causes confusion when referring to **molecular data** versus **morphological data**. The former term refers to characters that are solely molecules of some sort, which may be the nucleotides of DNA, the nucleotides of RNA, or the amino acids of proteins, for example. However, RNA and proteins are subject to biochemical actions such as post-translational and post-transcriptional processing, which can modify them from their original genotype form[12]. So, they are all correctly treated as part of the phenotype, not the genotype.

So, no matter what you may read, amino acids are not part of the genotype! For example, you did not inherit any amino acids from your father, only DNA. Morphological characters are part of the phenotype, and so are all molecular characters except those directly associated with the DNA. Of practical importance here, is the fact that an amino acid alignment represents phenotype, while a DNA alignment represents genotype.

*Phylogeny and Homology*

The term **phylogeny** refers to the shared history of the objects, whereas the term **homology** refers to the shared history of the characteristics (character states with a shared history are **homologues**). Inferred homology is used to group the objects into sets of descendants of some common ancestor, and inferred groups of objects are used to identify characters that have been inherited from that common ancestor. One can see this situation as being either circular (phylogeny ↔ homology), or as two sides of the same coin. Strictly speaking, we cannot treat amino acid alignment as the same topic as DNA alignment.

The objective of phylogenetics is to reconstruct both the phylogeny and the homology relationships. Phylogeneticists, as experts, may be interested primarily in the phylogeny, or in the homology, or in both. However, it is clear that non-experts (that is, those who are using a phylogenetic analysis solely as a tool for some other biological objective) are primarily interested in phylogeny. Therefore, far more computational research has been conducted on phylogeny reconstruction than on homology reconstruction (e.g. compare their respective treatments in Felsenstein[13] and Warnow[14]).

In practical terms, we usually represent the phylogeny via a fully connected line graph, where the nodes represent the objects (observed or inferred) and the lines represent their relationships. The lines are directed, so that the relationship is indicated from ancestor to descendant. It is thus a directed acyclic line graph (DAG).

We usually represent the homology via a table, in which the rows are the objects and the columns are the characters. The table cells indicate the state of the relevant character for the relevant object. When we are dealing with molecular data, this table is referred to as a sequence alignment. If there are only two objects then it is a **pairwise alignment**, whereas it is a **multiple alignment** for more than two sequences. It is obviously only the latter that is of interest to phylogeneticists, since the phylogenetic relationship between any two objects is trivial.

Note that the phylogeny (line graph) and the homology (table) are the two sides of the same coin referred to above — they are two representations of the same data. In an ideal world, we would be able to directly interchange information between these two representations. However, in the presence of ambiguous data this is not possible. So, in practice, we need to infer both the graph *and* the table.

Unfortunately, there is no simple conceptual or computational relationship between homology and phylogeny. Therefore, trying to address these two relationships simultaneously can potentially lead to serious errors. Phylogeny provides only part of the evidence for homology, and homology is not always necessary for estimating a phylogeny.

*Phylogenetic Homology*

Homology in the general sense simply means correspondence among the parts of a complex whole — we compare a set of parts and decide which ones correspond best. For example, I am bilaterally symmetrical, so my left hand corresponds to my right; but I am also a quadruped and so my left hand also corresponds to my left foot; and in my family I correspond to my sister, but I also correspond to my wife; etc. There are thus many meanings of "correspondence".

In phylogenetics, however, we mean a very *specific* type of correspondence, which differs in many ways from its meaning in other parts of biology, let alone other branches of science. Indeed, when Richard Owen originally defined homology in biology (way back in 1843), it was without any reference to evolutionary history.[15] It is a relatively modern idea to invoke evolution when discussing homology.

Of course, the fact that the concept of homology in biology has many different meanings confuses non-experts (and even some experts!). Morrison et al.[11] list eight different concepts of "homology" in biology:

Evolutionary homology, or phylogenetic homology
Character-state homology, or transformational homology
Character homology, or positional homology
Regional homology, or locus homology
Structural homology, or functional homology
Genic homology
Developmental homology, or deep homology
Organismal homology, or taxic homology.

Only the first of these is directly relevant for phylogenetic analysis. Nevertheless, the literature is replete with references to the word "homology" without any concern for distinguishing the various meanings; this is unnecessarily confusing. There is more discussion of these other meanings in Morrison et al.[11].

In phylogenetics, homology refers to characteristics that are shared because they have been inherited from a common ancestor. However, defining homology in terms of ancestry is neither logical nor operational. It is not logical because it refers to unique and unobservable historical events (historical accidents); and it is not operational because the relationship cannot be either observed or quantified.

This is the crux of the sequence alignment problem — we need to optimize a concept for which there is neither a direct biological quantification nor a simple mathematical representation.

Basically, we use comparative homology to try to identify phylogenetic homology.[16] That is, homology is inferred from the correspondence of parts (the characters) of a complex whole (the object), and we assume that both the object and its parts can be traced back to an ancestor. All we can do in practice is to try to adduce evidence for this ancestry, and evaluate the probability of that evidence, to produce the most likely hypothesis of homology for each character (and the phylogeny of the objects). This is an inference problem not really an optimization problem.

*Homology, Analogy and Similarity*

If we define homology is the correspondence between the parts of objects that are part of a historical lineage, then **analogy** is the correspondence between the parts of objects that are part of any other class (or set) of those objects.[17,18] In this sense, the objective of homology assessment is to distinguish homology from analogy. Analogy can arise from correspondence in the function of the parts, for example, or by stochastic processes. (Aside: analogy can be just as interesting as homology, depending on what you are studying.)

**Similarity,** or resemblance, can then be considered as the relationship between the wholes, not just their parts[17,18] — they are similar because a lot of their parts seem to correspond.

This distinction between similarity and homology is extremely important. For example, we may try to operationalize homology assessment by trying to maximize the 1:1 correspondences between the parts of the objects.[19] However, while this maximizes the similarity of the objects, it does not necessarily say anything about the homology of the parts. We may be optimizing analogy not homology.

Similarity is important for homology assessment, but there is more to it than this. This is an especially important point with regard to sequence alignment, as we shall see, because it has been based mostly on maximizing similarity alone — but similarity is not enough, because phylogenetics requires a **special similarity**: similarity due to homology alone.

**The Complexity of Homology**

I have tried to make it clear so far that this biological topic is not a simple one. What I need to do now is try to present the basic components of homology that seem to be most important for creating sequence alignments that address homology assessment. I will start with morphological characters, because that is conceptually the most straightforward. Only then will I address how the concepts relate to molecules.

Having observed a set of characters and their states for a set of objects, homology assessment consists of answering this question: Are these character states homologous for this set of objects? Sadly, there is no experiment that a biologist can conduct that would answer this question. Instead, the biologist needs to resort to some form pattern analysis of the dataset. This might be where bioinformaticians can make their contribution.

Let’s first look at an empirical example of homology assessment.

*Insect Bodies*

Reconstructing evolutionary history consists of constructing a scenario that covers (explains) the changes in the objects and their characters through their history of descent. Let's take the insect body as an example.

The body of contemporary insects consists of 3 segments: head, thorax, and abdomen (see Figure 1). The head has 3 sets of (paired) mouthparts, 1 pair of antennae, and 1 pair of compound eyes (plus some simple eyes). The thorax has 3 sets of (paired) legs, and (if present) 2 sets of (paired) wings. The abdomen has the respiratory organs, the digestive and excretory organs, and the sexual organs.

[Insert Figure 1.]

As one possible scenario (see Figure 1), we might conclude that, historically, the ancestor originally had one body segment, which has been duplicated twice through the evolutionary history of its descendants, and thus the three segments are homologous. That is, each of the three segments is a “whole”, and they have, or have had, similar “parts”.

Furthermore, in our scenario we might conclude that the ancestor had no appendages (the ancestral state), but that its descendants developed legs later on (the derived state). Even later, some of the legs were modified to function in feeding, not locomotion. This would mean that the three sets of mouthparts and the three pairs of legs are homologous. That is, each mouthpart set (the derived state) can be matched to a set of legs (now the ancestral state).

We might develop similar scenarios for the origin of the antennae and wings, which are not homologous with the legs, and which may or may not be homologous with each other.

Obviously, we can construct lots of such historical scenarios; and the more complex are the objects then the more scenarios we could have. For the biologist, deciding among the scenarios consists of evaluating as much evidence as can be gathered about the correspondences among the parts of the complex whole. Below, I will discuss the forms of evidence that can, and have been, used.

But first I need to discuss the hierarchical nature of homology.

*Homology and Hierarchy*

As noted above, defining homology in terms of ancestry is neither logical nor operational. It is not logical because homology is actually a hierarchical concept. This means that there is no simple answer to any question about homology — whether two states are homologous or not may depend very much on which level of the hierarchy the question is about.

The hierarchy arises from the fact that phylogenetic history has a strong hierarchical component — characters that arose early in history are now more widespread among species than are characters that arose later. The homology of some characters, therefore, occurs at a more general level than that of others. Indeed, features at lower levels in the hierarchy combine to generate features at higher levels.

Moreover, most characters are not inherited directly, and are thus part of the phenotype not the genotype. That is, asking whether two phenotypic character states (e.g. amino acids) are homologous is not meaningful, because we do not inherit them. If we don't inherit the characters, and homology is defined in terms of inheritance, how do we decide about their homology?

The classic example used to illustrate the hierarchical nature of homology (and also homology versus analogy) involves a comparison of the wings of birds, mammals and insects. It has long been accepted by biologists that mammals, birds, reptiles and amphibians are quadrupeds (i.e. they have four limbs) as far as their common ancestor is concerned. These limbs have been modified through evolutionary time from the lobe fins of certain fish-like organism. (Aside: fish have four fins for functional reasons, since they are needed for stability and movement in three dimensions.)

This is illustrated in Figure 2. This figure shows the relationships of the forelimbs ("front legs", called pectoral limbs) of birds and mammals. These limbs are assumed to have been derived from the lobe fins, so that the lobe fins are the ancestral state of these limbs. In turn, these limbs have been modified into a wide variety of ways, as various flippers, legs, wings and arms.

[Insert Figure 2.]

The organs in each row of the figure developed independently through time from the organ indicated in the row above. For example, the three types of flippers originated separately, as did the three types of wings, but they are all derived from pectoral limbs. The organs within the same row are thus homologous at the next upper level in the hierarchy, because they share the same ancestral organ. However, they are not homologous within their own level in the hierarchy, because they each arose independently from that ancestral organ. For example, all of the flippers and wings are **homologous as pectoral limbs** (i.e. they are homologues), but bird wings and bat wings are **analogous as wings** (i.e. not homologous).

So, there is no simple answer to the question: Are bird wings and bat wings homologous? The answer is both "yes" (as forelimbs) and "no" (as wings). The importance of this point is usually lost in all discussions of molecular data in relation to sequence alignment. Alignment of amino acids and alignment of nucleotides are not necessarily the same thing, because they exist at different levels of the homology hierarchy. This will be discussed in a later section.

For our example, it is also important to note that the wings of insects are not in any way homologous with either bird or bat wings. Insect wings are outgrowths of the exoskeleton of the thorax, rather than being modified legs. So, insect wings and bird/bat wings are analogous — they share a common function but not a common phylogenetic origin. That is, they are not homologous at any level of the hierarchy.

Finally, in relation to a hierarchy, there is the conundrum of "My Grandfather's Axe", or "The Ship of Theseus" — these are simply variants of the same conundrum. They refer to the situation where all of the parts of a complex object are replaced through time. For example, if each part of the ship gets replaced sequentially, the conundrum is: Is it still the same ship through time; and if not, when does it cease to be the same ship?

This is clearly a significant conceptual issue for phylogenetics, in which ancestral character states are replaced by derived states, often more than once. This can easily lead to situations where there are no obvious characters to compare, because they have all been replaced or lost. This becomes relevant when we compare characters at different levels of the hierarchy.

For our purposes, the characters of all organisms that share a common ancestor are homologous at *some* level of the hierarchy, no matter how much the character states have been modified, or even whether the character has been replaced entirely by something new. For example, snakes are quadrupeds, because they are descended from an ancestor that had four limbs. That is, snakes *do* have a homologue of forelimbs, but they have modified them so much that the character state is best coded as "absent".

*The Forms of Evidence for Homology*

A phylogenetic correspondence (i.e. a match) occurs only between parts that have a historical relationship, in terms of being descendants of the same ancestral part. We cannot *know* about this, but we can devise the most probable relationships among the elements given whatever information is available to us. In the presence of different evidence we might come to a different conclusion; and with increased evidence we might update our hypothesis or replace it.

Evidence comes from considering the levels of the hierarchy. This is discussed in more detail by Morrison[20] and Morrison et al.[11] In summary, Patterson[21] identified several tests for making decisions about potential homologies in an objective and repeatable manner. However, for our purposes here, it is better to treat them as criteria to be evaluated, rather than as tests to be decided. These are listed in Table 1.

**Table 1.**

Criteria developed by phylogeneticists to evaluate potential homologous relationships among the parts of a set of complex wholes. The criteria are applied when comparing the parts between different wholes.

| CRITERION FOR HOMOLOGY | DESCRIPTION |
|---|---|
| Similarity | |
| Compositional | apparent likeness or resemblance of the parts |
| Ontogenetic | correspondence of the developmental sequence of the parts |
| Topographical | correspondence of the three-dimensional locations of the parts |
| Functional | correspondence of function among the parts |
| Conjunction | character states can be homologous only within a single character |
| Congruence | agreement with other postulated homologies of other characters (i.e. synapomorphy in a phylogenetic tree) |

There are three basic criteria (**Similarity**, **Conjunction**, **Congruence**), but there are four types of Similarity, which can be evaluated independently. It has been suggested[21,22] that hypotheses of homology initially come from noting Similarity, and that they are then tested using Congruence. In this view, the initial alignment table (objects arranged horizontally, characters arranged vertically) would be created based on maximizing Similarity, and Congruence would then be tested by constructing a phylogenetic tree.

Conjunction (or coexistence), on the other hand, refers to some sort of serial homology among multiple copies of the same part, as among the insect body segments referred to above. Homology among the body parts, due to their two duplications through evolutionary time, is not the same thing as homology due to modification of a single part. This issue of homology among repeated parts will be important when we come to discuss molecular data.

The basic problem for homology assessment is that character states that are apparently identical are not necessarily homologous, let alone ones that merely appear to be somewhat similar. For example, the observation of uni-locular ovaries in a set of plant species does not make these ovaries homologues. One subset of ovaries might develop from their primordia as uni-locular, while another subset may develop as bi-locular and then lose the dividing septum. The end result will look the same. This sort of scenario involves the evolutionary processes of parallelisms and convergences, which result in analogues not homologues.[23]

So, simple similarity among the parts of a set of complex wholes cannot be enough for proposing hypotheses of homology. It is for this reason that four different types of similarity are used in biology: **Compositional**, **Topographical**, **Ontogenetic**, and **Functional**. That is, we might see homology assessment as referring to a "special similarity", in that apparent resemblance should ideally involve all four similarity criteria, simultaneously. Any one criterion on its own can be misleading, and so their combination will provide the best evidence we can get.

In the ovary example that I just used, the ovaries have compositional similarity (they look the same), topographical similarity (they are located in the same part of the flower) and functional similarity (they contain the ovules, which will develop into seeds), but they do not have ontogenetic similarity (they develop from their primordia in different ways); and so they are not likely to be homologues. It also turns out, in the specific example I have in mind (the genus *Goodenia*), that they are also not congruent on a phylogenetic tree, which confirms their lack of homology. The two subsets of ovaries should be treated as different characters.

Sadly, the ideal goal that we will detect homology because all four types of similarity plus conjunction plus congruence will agree with each other does not often occur in practice. This means that homology assessment involves balancing the different criteria, when they disagree. What is the most likely / probable hypothesis, given the evidence? This is the crux of the computational problem that needs to be solved in order to provide an algorithmic homology assessment.

Finally, an alignment represents the historical events that have occurred, irrespective of whether it refers to phenotype (morphological or molecular) or genotype. The alignment is thus a static representation of a dynamic set of processes. This is ultimately what causes all of the representational problems, because there is no necessary and sufficient way to achieve this representation.

## Molecular Data

We are now in a position (at last) to consider how the conceptual treatment outlined above applies to molecular data, so that we can produce a multiple sequence alignment that maximizes homology. This topic is discussed in more detail by Morrison[20] and Morrison et al.[11]

### *Molecular Alignments*

As noted above, multiple sequence alignments are simply a tabular arrangement of the objects and their characters. The rows are the objects (usually an individual organism) from which the molecular data have been collected. The columns are the characters, which may be nucleotide positions in a DNA sequence, amino acid positions in a protein sequence, AFLP primers, restriction enzymes, etc. For linearly ordered characters, such as nucleotides and amino acids, it is conventional to arrange the characters in the order in which they are "read" by molecular processes. However, this is not strictly necessary; and, as we shall see, it is not always practical. Most other molecular characters can be arranged in any order at all.

The cells of the table are the character states actually observed. These will be nucleotides, amino acids, presence/absence scores, etc. It is important to note that the objects and the character states are experimentally observed, but the characters are not necessarily observed. This is actually at the heart of the need for a method of sequence alignment.

For molecular data involving AFLP primers and restriction enzymes, for example, the characters are pre-determined by the process of collecting the data, and so which character state goes with which character is obvious. However, for data such as DNA or protein sequences, and also for morphological data, the number of characters is, in practice, determined by the data analysis procedure.

In this regard, Morrison et al.[11] clearly distinguish between the unit of observed variation and the inferred character state. For nucleotides, there are only four units, represented by the symbols A, C, G and T (or U, if we are dealing with RNA). All As are identical, and so are all Cs, etc. However, the nucleotides are not intrinsically character states. A nucleotide unit becomes a character state only when it is assigned to a particular character. If it is assigned to a different character, then it is a different state, even though it is still the same nucleotide (unit).

The alignment process then effectively consists of shuffling the units among the various possible characters (but not the rows), or creating new characters if needed, to see which column each observed character state "best fits into", and thus deciding which character states they are. This is a computational problem that should be based on maximizing the character-state homologies.

Sadly, the conceptual ideas outlined in the previous section are not necessarily easy to apply to molecular characters. For most such characters, there simply is not much information from which to evaluate the homologies.

Take nucleotides, as the most obvious example. What evidence do we have to put any particular A, say, in any particular column? That is, which character does each unit belong to? Or, equivalently, what is its character state? Two As represent the same character state only if they originated as As as a result of the same evolutionary event (i.e. an A is only the same as an A with which it is aligned).

This problem often does not arise when dealing with morphological data. The parts of the complex wholes, which are being compared, are themselves often quite complex. This potentially makes the evaluation of the various forms of similarity somewhat straightforward. An insect leg tends to look like an insect leg because of its complexity (i.e. they have compositional, topographical, ontogenetic and functional similarity); and only when we try to compare legs to mouthparts does the complexity potentially create problems.

The same cannot be said of nucleotides (or amino acids). Their direct comparison is almost meaningless. We need more information, such as their linear order, before we can hope to make progress. So, it is the much vaunted simplicity of molecular characters that actually creates the alignment problem. Far from being "better" than morphological data, molecular data are actually harder to deal with.

Nevertheless, we must proceed, anyway. NP-hard problems have never stopped computationalists before!

*Homology of Molecular Data*

First, we need to remember that, for molecules, the only genotype elements are nucleotides within a chromosomal locus, and genes within a genome. This is the "bottom" level of the homology hierarchy. Everything else is part of the phenotype, and forms the various hierarchical levels above. One possible example of this molecular hierarchy, involving protein-coding genes, is illustrated in Figure 3.

[Insert Figure 3.]

This figure illustrates different levels of biological organization. The arrows in the figure show the functional dependence of some of the different types of molecular data, and how they control the development of organ systems. The study of each of these levels of molecular complexity has its own theoretical and practical goals. Consequently, their functional dependence does not invariably imply evolutionary homology at each step in the process.

Nucleotides can be homologous with one another because of shared inheritance from a nucleotide in a common ancestor (i.e. they actually *are* copies of that ancestral nucleotide). However, all of the other data are phenotypic, because they are expressions of the inherited genotypic data, rather than being directly inherited themselves. Therefore, we cannot automatically infer that homology at any one step in the dependence chain leads to homology at any other step.

In particular, character-state substitutions at any historical step may break the implication of homology at any other step. For example, one protein domain of an enzyme may be substituted for another during evolution, and yet the enzyme can still function to catalyze the same biochemical pathway, and thus lead to the development of the same organ. In this scenario, the organs would still be homologous between species, even though the domains are not. Therefore, if we consider two organs to be homologous, that does not mean that we can infer that all of the associated enzyme domains are homologous; and nor necessarily are two organs homologous just because we consider them to share all of their enzyme domains.

A similar situation applies to all of the other relationships among molecular data types. For example, substitution of one nucleotide for another does not mean that the coded amino acids are no longer homologous; and homology of amino acids does not mean that all of their associated nucleotides must be homologous.

[Insert Figure 4.]

A simple example is shown in the multiple sequence alignment of Figure 4. There are 15 nucleotide positions (characters) shown for each species, which code for a sequence of 6 amino acids (the phenotype). The decisions about the homology of the nucleotide character states is unproblematic for most of the alignment positions, based on both compositional and topographical similarity (see the next section). However, in the first sequence, ontogenetic similarity suggests a different scenario. It is likely that the molecular process known as slipped strand mispairing (or replication slippage) has caused the duplication of a T (a tandem repeat), followed by a compensatory deletion of an A. This deletion has a functional purpose, as it maintain the rest of the translational reading frame (i.e. only two of the amino acids will not be correctly coded). In this scenario, the nucleotides are homologous within the columns as shown in the figure, with the extra T requiring the recognition of an extra character, compared to the other sequences, with the character states coded as "not applicable" for those other sequences.

In this scenario, the amino acids remain homologous within each of their 6 columns, in spite of the fact that some of the nucleotides code for different amino acids in different rows — the amino acids could easily be displayed as an ungapped multiple alignment. So, some of the nucleotides have changed their function through time (they have different codon positions in the amino acids), but they are still inherited from a common ancestor, and so their homology is maintained. The nucleotides are homologous and the amino acids are homologous.

However, the homology relations of the amino acids and the nucleotide *cannot* both be displayed within the same alignment. This is an inevitable consequence of the hierarchical nature of homology. It will not be possible to show homology relationships across more than one level of the hierarchy, unless the homology relationships are also maintained across those levels. This is an important practical point for the output from an alignment computer program.

*Evidence for Molecular Homology*

Obviously, one of the main points of this paper is that the same evidence that phylogeneticists have always used for assessing homology should be applied to molecular data, as well. This is not a point that has often been made, mainly because compositional similarity has played such a dominant role — the other criteria have effectively been ignored in the main stream of algorithm development. Morrison[20] discusses this topic in more detail.

**Table 2.**

Evidence used for the various homology criteria when applied to nucleotide sequences.

| CRITERION FOR HOMOLOGY | DESCRIPTION |
|---|---|
| Similarity | |
| Compositional | % sequence similarity |
| Ontogenetic | inferred molecular mechanism creating the sequence variation |
| Topographical | second- and third-order structure of the coded protein or RNA |
| Functional | annotated function of the sequence in the coded protein or RNA |
| Conjunction | within-genome copies of the same sequence (i.e. paralogy) |
| Congruence | agreement with other postulated homologies elsewhere in the same sequences (synapomorphy) |

Table 2 lists the way in which the different evidence types relate specifically to nucleotide sequences. At the outset, it is worth noting that these pieces of evidence can be evaluated independently, and that they may produce different alignments. That is, the evidence may point in different directions. To illustrate this point, Morrison[20] provides two figures (his Fig. 2 and Fig 3) that provide homology hypotheses based on the different types of similarity. These alignments all differ to one extent or another.

We can now take each evidential feature one at a time, and look in some detail at what it means in terms of data analysis. In each case, the discussion refers to the comparison of two or more nucleotide sequences, to assess which nucleotides are homologous. How are the criteria to be applied?

**Compositional similarity** refers to the identity of nucleotides when arranged in their sequence order — this is usually the intended meaning when the expression "sequence similarity" is used. Identity is converted to similarity based on some substitution scoring scheme plus the penalty given to indels (see below). The scoring scheme specifies how non-identity of nucleotides should be scored, based on some notion of how "easy" it is for any given nucleotide to be substituted by any other nucleotide during evolutionary history, or to be inserted or deleted.

The logic of applying sequence similarity to homology assessment is straightforward:

$$\text{similarity} = \text{homology} + \text{analogy}$$
$$\text{if analogy} \rightarrow 0$$
$$\text{then similarity} \rightarrow \text{homology}$$

Similarity can then be used as a surrogate for homology relationships. For this reason, in molecular biology "homology" is often treated as a synonym of "similarity".[24] This idea dates back at least to Margoliash[25]; and it continues to this day when database searching is referred to as "homology searching",[26,27] molecular structure prediction is referred to a "homology modelling".[28]

It is unfortunate that the word homology has long been used as a synonym for similarity.[29,30] Similarity is one of several criteria that can be used to help infer homology, but making the words synonymous confuses empirical measurements (similarity) with inferred conclusions (homology).

Furthermore, analogy is only small (analogy → 0) when the sequences are very similar, and so for most of the identity range (0–100%) analogy is not trivial; indeed, below 25% identity it may dominate homology. Morrison[3] has a graph (his Fig. 4) which shows that nucleotide sequence identity >80% is required before analogy becomes unimportant, for practical purposes (e.g. >90% success at identifying homology); this is equivalent to a distance of c. 0.5 substitutions per site when using the HKY substitution scoring scheme.[20]

What is more problematic, however, is that the measurement of sequence similarity assumes that substitutions and indels occur at random.[31,32] However, sequence variation occurs distinctly non-randomly and non-independently, both in space and time, due to the many molecular mechanisms causing sequence variation (discussed below, under ontogenetic similarity), so that sequence variation is anything but random even locally.[11]

Thus, compositional similarity only works well as a criterion in highly conserved sequence regions. For example, compositional similarity fails as a criterion when there are closely adjacent but independent insertions / deletions, producing what has been termed "over alignment".[33] In fact, this non-randomness actually contains valuable information about homologies. It is for this reason that we cannot ignore the other types of similarity: ontogenetic, topographical and functional. It will be most effective to discuss ontogenetic similarity first.

**Ontogenetic similarity** refers to the use of developmental processes and timing as criteria for recognizing homology. For nucleotide sequences, this will involve identifying which of the known molecular processes is most likely to have caused the observed sequence variation. The multiple alignment would then represent a set of scenarios for how the set of sequences varied through evolutionary time.

The ontogenetic criterion makes it clear that homology refers to the relationship between parts of organisms that have resulted from the same heritable transformation events. These events should be used in creating the alignment rather than being a posteriori deductions from it. This is discussed in more detail by Morrison.[34] The recognition of the importance of using this information in multiple sequence alignment goes back 25 years or so,[35-37] although it was actually first used in 1983.[38]

**Table 3.**

Molecular mechanisms that can lead to nucleotide sequence variation, along with their associated observations and models.

| MOLECULAR MECHANISMS | CURRENT MODELS | OBSERVATIONS |
|---|---|---|
| Substitutions | substitutions | no length variation |
| Inversions | substitutions | no length variation |
| Transpositions | substitutions | no length variation |
| Duplications | | |
| Tandem repeats | indels | length variation (gaps) = not applicable |
| Inverted repeats | indels | length variation (gaps) = not applicable |
| Translocations | | |
| within studied sequence | substitutions | no length variation |
| outside sequence | indels | length variation (gaps) = not applicable |
| Deletions | indels | length variation (gaps) |
| Insertions | indels | length variation (gaps) = not applicable |

The key to using this criterion is recognizing that the observed sequence variation results from known mechanisms, and that we need computational models to connect the observations and the mechanisms, as indicated in Table 3.

The basic observation is that the sequences being compared either vary in length or they do not. When using sequence similarity alone, all length-invariable differences among the sequences are currently modelled as substitutions, and all length-variable differences are modelled as indels (insertions or deletions). These are very simplistic models, given that sequence variation is caused by a wide range of molecular mechanisms, including:[34]

substitution (replacement of one nucleotide by another);
inversion (replacement of a subsequence by its reverse-complement);
transposition (exchange of subsequences between locations);
translocation (removal of a subsequence and its insertion at another location);
duplication (copying a subsequence), notably tandem repeats (copying to an immediately adjacent position) and inverted repeats (reverse-complementing the copy);
insertion (addition of a novel subsequence); and
deletion (removal of an existing subsequence).

These mechanisms are controlled by various cellular processes, including polymerase error, recombination, transposable elements, and DNA repair mechanisms. Of these processes, tandem repeats are probably the most common cause of sequence variation.[39]

An empirical example of a multiple sequence alignment based on the ontogenetic criterion is shown in Figure 5. Seven independent events in the evolution of the variation among the 24 sequences are numbered, and the subsequences involved are boxed. Events 1, 5, 6 and 7 all involve tandem repeats of short subsequences (presumably resulting from slipped strand mispairing). Events 2 and 4 involve deletion of short subsequences (possibly also by slipped strand mispairing). Event 3 is likely to be an inversion, as it occurs in the loop region in a stem-loop secondary structure. All of the other events required for the alignment are substitutions, except for a 1-nucleotide deletion at position 47 (which might even be a sequencing error).

[Insert Figure 5.]

Perhaps the most important points to note about these mechanisms are that: (i) they often occur non-randomly across sequences; (ii) many of them affect multiple nucleotides as a single unit (e.g. duplications, inversions, translocations, transpositions); and (iii) some of them change the order of the nucleotides relative to the other sequences (e.g. inversions, translocations, transpositions).

It is the combination of points (i) and (ii) that results in simple sequence similarity failing to detect homology. The current computational models treat both substitutions and indels as independent and identically distributed (IID) random variables. They thus fail to model quite a few of the molecular mechanisms. For example, all sequence mismatches are modelled as IID substitutions (e.g. a 4-nucleotide inversion is modelled as four independent substitutions) and all length variations are modelled as IID indels (e.g. a 6-nucleotide tandem repeat is modelled using an affine cost for a variable-length indel). Sequences violate the assumptions of these models far too much for those models to produce multiple sequence alignments that can stand up to even casual scrutiny. This same issue affects all attempts to evaluate alignment algorithms by using simulated sequences,[40] as all of the simulations programs also make the IID assumption.

The current models also assume that all of the sequences are co-linear, which is violated by point (iii). This issue is not addressed by most multiple sequence alignment algorithms. It is, however, addressed by most programs designed for gene alignment,[41,42] since it is widely recognized that gene order can vary widely among genomes.

**Topographical similarity** refers to the topological correspondence between the parts of the whole. That is, it uses relationships of the parts *within* the same sequence to identify possible homologies, whereas the other forms of similarity look for relationships *between* sequences. For nucleotide sequences, this similarity is based on the second-order (planar) and third-order (three-dimensional) structure of the encoded gene product (as described by Morrison[34]).

For example, for RNA-coding sequences (and non-coding sequences such as group I & II introns, and transcribed spacers), the most important second-order structures are the stem-loops, where some of the RNA nucleotides are paired (the stems) and some are not (the loops). Similarly, for protein-coding sequences, codons make up the most important second-order structural feature. Third-order structures for both RNA-coding and protein-coding sequences consist of long-range interactions between parts of the structures.

These structures place non-random biological constraints on the coded macromolecule, and different regions of the sequence will have different functional constraints. Where stem-pairing or codons enforce constraints, compositional similarity of the original nucleotides will not necessarily be maintained through evolutionary time. However, topographical similarity will usually still be maintained — changes in the nucleotides are tolerated by the molecular processes provided that the RNA nucleotides can still pair, or the required codons are still coded. Topographical similarity can then still be an effective criterion for recognizing homology, even when compositional similarity fails.

As an example, Morrison[3] reports a 429-nucleotide alignment of 10 sequences, in which the nucleotides show only 66% identity, and yet there is no variation whatsoever in the coded amino acid sequences. That is, the topographical similarity is perfectly maintained, while the compositional similarity is not.

An RNA-coding example is shown in Figure 5, where part of the sequence forms a stem-loop structure. The nucleotides in the 5' part of the stem pair with their reverse-complements in the 3' part of the stem, and those nucleotides that are involved in corresponding stem-pairs have been aligned. Events 1 and 5 each create a bulge in the stem-loop structure, while Events 2 and 4 disrupt the pairing of that structure. The other Events do not affect the functioning of the structure.

There are limits to the use of topographical similarity, of course. For example, for RNA-coding genes, topology is not conserved in the so-called regions of ambiguous alignment or regions of expansion and contraction.[43] Nevertheless, topographical similarity is often proposed as a criterion for assessing the quality of multiple sequence alignments generated using other criteria.[44]

**Functional similarity** is often treated as a separate criterion when dealing with morphological data. For example, Richard Owen's original definition of homology referred to identity of organs irrespective of form or function, which appears to separate structure and function (and Owen apparently meant homology = similarity of structure, whereas analogy = similarity of function).

However, for molecular data, function is closely associated with topographical similarity, because molecular structure and function are usually so intimately related. For example, it is recognized that bulges in RNA stem-loop structures are involved the functioning of the RNA, while the paired stems function to hold the bulge in the 3-dimensional location required for that function.[34]

Unfortunately, function is not necessarily well-defined in a set of sequences, as single nucleotides may have multiple functions (e.g. be involved in both second-order and third-order structures), and multiple nucleotides may share a single function (usually a second-order structure). This means that sharing the same function may be evidence of homology between nucleotides, but having different functions is not necessarily evidence of analogy.

To quote Fitch[45] (p.231): "Life would have been simple if phylogenetic homology necessarily implied structural homology or either of them had necessarily implied functional homology. However, they map onto each other imperfectly."

**Conjunction** addresses the issue of repeated relationships within sequences. Strictly speaking, we cannot have homologues *within* a sequence if we are making comparisons *between* sequences. That is, the coexistence of homologous features as multiple copies within the same sequence "muddies the waters" for sequence alignment. If a subsequence is repeated in one sequence, then there is no way to determine which of the copies is homologous to the single copy in some other sequence.

For example, for the tandem repeats shown in Figure 5, should the first or the second copy be aligned with the single copies? I adopted the simple convention of aligning the first copy, on the basis that the repeats arise from slipped strand mispairing, which would make the second one the copy; but this is only a convention.

This issue is important for molecular data, because gene products are usually expressed at levels of more than one copy per cell, and there are many greatly repetitive genes (such as rRNA and histones), so a rigid interpretation of conjunction is not practical. Molecular biologists have addressed this issue by defining a series of types of homology. **Orthology** is the homology between nucleotides in different sequences arising from common descent. **Paralogy** is the homology between nucleotides within the same sequence (e.g. repeated subsequences). **Xenology** is homology due to so-called horizontal phylogenetic processes (e.g. hybridization, horizontal gene transfer), rather than vertical inheritance by descent.

For the purposes of multiple sequence alignment, the products of all three types of homology can provide phylogenetic evidence. However, only orthologous nucleotides should be aligned in a multiple sequence alignment. Nucleotide relationships arising from paralogy or xenology should not be aligned, because they do not meet the phylogenetic definition of homology that we require.

The important point here is to not align (in the same column) features that are not orthologous, so that unaligned nucleotides are clearly indicated as non-orthologous. This requires the addition of extra characters to accommodate paralogy and xenology. The aligned position in the other sequences should then correctly be treated as "not applicable" (i.e. there is no homologue), just as they should be for any insertion.

This potentially causes confusion. Current models do not distinguish between different types of gaps in a multiple sequence alignment. Following Table 3, a gap is an observation — we observe sequences of different length in the alignment. However, this is hard to model, and so the gaps are all modelled as indels, even though some of the gaps refer to deletions of subsequences (the homologues are now gone) and others refer to various insertions (there are no homologues and never were). This treats both "absent" and "not applicable" as the same character state, as recognized by Löytynoja and Goldman.[46]

For this reason, some practitioners delete gapped regions as being the most likely locations of non-homology, presumably leaving only those regions well-aligned by compositional similarity. Other practitioners manually adjust alignments to rectify the most obvious failures of compositional similarity, in terms of identifying phylogenetic homology.

**Congruence** recognizes that congruent patterns among multiple postulated homologies provide strong evidence that the inferred homologies are correct. It is for this reason that some people view homology as being inferred from similarity (called primary homology) and tested by congruence (called secondary homology), because congruence is the only criterion directly tied to ancestry.

Using the example of bird wings and bat wings discussed above, treating them as homologous does not pass the congruence test because they do not form a synapomorphy on a phylogenetic tree produced from other morphological data. This implies that there these wings had separate evolutionary origins.

There is no substantive difference between the congruence criterion as applied to molecular or morphological data. Its basic problem is that congruence cannot be the sole arbiter of homology, because it treats synapomorphy (= taxic homology) and phylogenetic homology as being synonymous. Mere congruence of characters alone cannot identify homology— homology implies synapomorphy, but apparent synapomorphy does not necessarily imply homology. All we have is an *estimate* of the true phylogeny, which will be inaccurate to one extent or another, and so congruence is unlikely to be a successful criterion for detecting homologies on its own.

One well known example of this situation is so-called long-branch attraction, often associated with sequence data.[47] Here, spurious compositional similarity of character states among sequences of rapidly evolving taxa overcomes the expectation that apparent synapomorphy is most likely to be equivalent to phylogenetic homology.

## Computations and Sequence Alignment

I am now in a position to relate the molecular concepts outlined in the previous section with the computational problems involved in trying to make the concepts operational (in spite of the fact that homology relationships cannot be either observed or quantified).

### *What has been done in the Past?*

I will start this section with a brief history of multiple sequence alignment. This history is rarely seen as important,[1,2] but presenting it serves an important rhetorical purpose. That is, I am doing this **not** because history is necessarily interesting, but because it clearly illustrates what biologists tried to do before they had computers to help. This will aid us in understanding what the biological goal should be. What did the biologists do to assess the homology relationships among molecular sequences, when left to their own devices? What sorts of sequence alignments did they produce, and how? I can tell you now that they differed considerably from what is produced by current alignment programs.

To a bioinformatician, the history of sequence alignment starts in 1970, with the presentation of the dynamic programming algorithm of Needleman and Wunsch.[48] However, protein sequencing started fully 20 years earlier than this,[49] and by the end of the 1950s comparisons of amino-acid sequences among related organisms were beginning to appear. However, as noted by Eck:[50] "data on amino acid sequences can be sorted, tabulated and arranged in a great variety of ways ... Any such manipulation will produce some sort of pattern." That is, analogy can be just as interesting as homology, depending on what you are studying. Thus, what is now called a multiple sequence alignment was only one of many possible data presentations, and not necessarily the most obvious one unless intended for an evolutionary analysis (e.g. phylogenetics).

For example, most of these early comparative studies focused on the structure (and thus function) of the proteins rather than on their evolution, and so they tended to present visual juxtapositions consisting of ungapped fragments of the sequences,[51-53] particularly the functionally active regions. Other studies were directed towards finding a solution to the problem of the genetic code (i.e. how nucleotides code for amino acids), and their presentation of sequence alignments was similarly non-evolutionary.[54,55]

Nevertheless, the early work on molecular evolution did reveal that different protein molecules are homologous, including what are now called paralogues.[56,57] With the sequencing of those proteins, it soon occurred to several people independently that the relative *positions* in the amino acid sequences are homologous, as well.[58] This is an important distinction, because the latter refers to the 1:1 matching of the parts (amino acids) of a complex whole (the protein molecule). However, most sequences were still presented unaligned,[57] until the work of Margoliash[25] and Pauling and Zuckerkandl.[59]

The major problem with sequencing proteins was that in the 1960s it was still a slow and tedious procedure, so that data were rather scarce — the first major compilation of aligned sequences appeared only in 1965.[60] Strasser[61] provides an interesting coverage of the early uses of multiple amino-acid sequence alignments, including the development of one-letter codes for each of the amino acids in order to make the alignments more readable. García-Sancho[49] and Suárez-Díaz[62] discuss the subsequent development of methods for the sequencing of RNA in the mid-1960s and, finally, DNA in the mid-1970s, which greatly increased the practical need for automated sequence alignment.

Most importantly, a number of the early molecular sequence alignments were constructed by hand, explicitly based on evaluation of the likely biological mechanisms that had produced the sequence variation. That is, the alignments make clear the originating molecular mechanisms. For example, Pauling and Zuckerkandl[59] provided a pairwise alignment of two reconstructed ancestral amino-acid sequences of haemoglobin along with a discussion of the substitutions and insertions/deletions involved.

Twenty years later, in what appears to be the first published study of intraspecific variation using DNA sequences, Kreitman[38] took this idea further, and provided a very carefully considered multiple alignment based on explicit recognition of tandem repeats and RNA stem structures within the study gene. This was very much in line with traditional approaches to the assessment of homologies (prior to phylogenetic-tree building), for example when using phenotypic characters.

However, shortly afterwards practical computerized procedures were developed based on dynamic programming for pairwise alignment (based solely on maximizing compositional similarity — note the title of the paper by Needleman and Wunsch[48]) and combined with the progressive alignment strategy for the multiple alignment step.[63] Then the CLUSTAL computer program was released, which implemented these procedures in a usable manner for personal computers,[64] and the history of the study of molecular evolution was changed forever.

CLUSTAL has come a long way since then.[65] These days, many more programs exist,[7-9] with different optimization criteria, different algorithms and heuristics, different scalability, and indeed sometimes very different philosophies. All of this effort has led to a proliferation of available alignment methods, which produce detectably different multiple sequence alignments in almost all realistic cases.

We learn four important things from this brief history:

(1) Early studies of molecular data were not directed towards phylogenetics, even when sequence alignment was involved. There is more to the study of evolution than phylogenetics, and much more to the study of biology than evolution. A multiple alignment can present data for many purposes.

(2) Sequence alignment originally involved amino acid sequences, not nucleotide sequences, and thus referred to phenotype not genotype. Homology assessment was, in this sense, no different from that previously applied to morphological data.

(3) Early attempts at multiple sequence alignment involved explicitly looking for evidence of homology, such as processes of molecular change and second-order molecular structure, thus using ontogenetic and topographical similarity as criteria for homology.

(4) Early computer programs were based solely on compositional similarity (of the whole) rather than homology (of the parts). This is presumably because algorithms already existed, in terms of computational edit processes.

The bottom line is that the biological goal is quite different to the current computational one. We need to change the latter (not the former!).

*Algorithms for Different Criteria*

Having completed my discussion of the biological goals of multiple sequence alignment, I have achieved the substantive objective of this paper. However, obviously I should say something about the algorithms that have already been developed that could be used to address the topic computationally.

The basic objective of using these algorithms is that homologies need to be "discovered" within the molecular data. This is the distinction between the ontological definition of homology (characters sharing common ancestry) and the epistemological diagnosis of homology (some sort of observed shared similarity). We need to make homology analysis operational, if we are to discover homologies.

Algorithmically, sequence alignment has traditionally been seen as a string-matching procedure.[66,67] So, multiple sequence alignment has been computationally about minimizing the edit distance between sequences. The strings are edited in specified ways, so that the final arrangement optimizes some mathematical objective function measuring the matches and mismatches.

This approach is thus based on compositional similarity as criteria — it maximizes the pairwise similarity of the sequences, often using dynamic programming as an exact algorithm. The objective function being maximized is based on nucleotide identity, or defined with respect to a substitution matrix (a log-odds scoring scheme), and a penalty for introducing indels into the sequences. This approach has generally been called “optimal alignment” in the bioinformatics literature; and there are plenty of programs using it (e.g. CLUSTAL, DIALIGN, MAFFT, Muscle, Probcons, T-Coffee).

It has even been suggested that all alignments should be possible ab initio, using only the (compositional) information in the sequences themselves,[68] and to this end much of the algorithmic sophistication of recent programs has been directed at compositional similarity alone (sometimes then called positional homology).[69-74]

However, it has long been known that even an optimal pairwise sequence alignment is not necessarily the homologous alignment.[3,75] That is, compositional similarity is not on its own a viable criterion for homology. Therefore, optimal alignment has probably been the biggest hindrance to the development of an automated homology method, as it has distracted both the theory and practice away from the main biological issues into computational ones that have shown little promise for dealing with homology assessment. A better discussion would be about what things should be included as “edits”. Operationally, sequence alignment consists of evaluating the probability of the nucleotides being a character state of each of the available characters.

Most approaches to multiple sequence alignment have relied on a two-step algorithm: (i) maximize the pairwise relationship of the sequences; and then (ii) maximize the relationship for multiple sequences. So, we could start the discussion by looking at algorithms that exist for comparing sequences pairwise, based on the other criteria I have outlined: topographical, ontogenetic and functional similarity, plus conjunction and congruence.

Morrison[20] provides an introduction to the primary literature, and so I will only briefly summarize the topic here.

There are currently no general computer programs based directly on topographical similarity as a criterion for multiple alignment. There are, however, programs that will do this indirectly, for example, by translating nucleotide sequences to amino acid sequences and then using compositional or topographical similarity for alignment of the amino acid sequences; but there are objective criteria by which the success of these alignments can be judged. There are also programs that will attempt to align RNA-coding sequences based on their inferred secondary structure as well as tertiary structure, but it is clear that they are currently not really adequate for the data sets used in phylogenetics;[76] and so this criterion is mostly implemented manually.

Several published papers have provided suitable objective criteria for manually creating these RNA alignments. Morrison et al. [11] discuss the pros and cons of manual versus automated alignment procedures. Furthermore, there are now many databases of sequences aligned according to their secondary structure.[77]

There are currently no computer programs for global multiple alignment of nucleotides based directly on ontogenetic similarity. However, work has been done for single sequences involving repeats and inverted repeats; and for pairwise alignments involving duplications, inversions, rearrangements , and their combination. However, in most cases there is no algorithm that clearly outperforms the others.[78] There are also programs based on multiple local alignments of deleted, repeated or rearranged sequence blocks but not individual nucleotides.

The use of profiles or hidden markov models of so-called intermediate sequences as templates for the pairwise alignment step[79] can be seen as a heuristic attempt to incorporate some of the information from ontogeny into multiple alignment. However, to date, this criterion has mainly been implemented manually for DNA sequences; and several published papers have provided suitable objective criteria for creating the alignments.

Functional alignments have been based mostly on the finding of sequence motifs associated with experimentally annotated functions. There are many specialist programs for this, although they search only for very specific motifs,[80] and so cannot be generally useful.

Conjunction appears to have never been investigated as a separate issue algorithmically, in the sense that any duplications will violate the assumptions of conjunction, and so the issue arises directly from the study of repeated sequence blocks.

So-called “phylogeny aware” alignment algorithms (e.g. SATé) are a heuristic attempt to use congruence as a criterion for nucleotide alignment, but their limitation is the requirement for a known phylogeny. Alternatively, there are currently two computerized approaches that operate by combining alignment and tree-building into a single procedure: statistical alignment (e.g. BAli-Phy, StatAlign) and direct optimization (e.g. POY, MSAM, BeeTLe). In essence, both of these latter approaches build trees directly, using models that include indels in addition to substitutions. The alignment is then an “implication” from the tree itself, rather than being derived separately. The basic issue is that this approach makes congruence the main criterion for detecting homology, by trying to combine compositional similarity and congruence. The duality between alignment and tree building does not mean that they must be inextricably confounded.

This issue is related to the oft-repeated concept that simple models may actually be better than complex ones, for any particular data-analysis purpose. A complex model may well result in the well-known failure of “not seeing the woodland for the trees”. To this end, if the objective is a phylogeny rather than alignment, then accurately identifying the homologies may not actually be necessary.

It has also been observed that complex nucleotide substitution models may also not be necessary even to create an “accurate” multiple sequence alignment.[81] I therefore emphasize the role of sequence alignments in representing homology — this is a valuable role in and of itself. It is entirely possible to adopt the opposite approach, and make alignment the sole objective rather than phylogeny. Indeed, Jardine[19,82] tried to do precisely this, alignment without phylogeny; and his algorithmic work is worth a second look in the modern world.

*Constructing an Algorithm*

In this section I will **not** present much in the way of computational algorithms. Instead, I will focus on understanding which aspects of sequence alignment are important in determining the biological goals of the computations.

Perhaps the most important point is that a multiple sequence alignment can be a goal in itself, when used for phylogenetic purposes — that is, when it represents homologies. Morrison et al.[11] list five possible uses of an a multiple alignment that are independent of any phylogeny, and a further four aspects of phylogeny-based biological estimates that can depend critically on the alignment. A “phylogenetic alignment” is thus much more than just a tool.

It seems rarely to be appreciated that a sequence alignment potentially contains more evolutionary information than does a phylogeny. The phylogenetic graph is simply a diagrammatic summary of *some* of the tabular information contained in the alignment. That is, several alignments may imply a single phylogeny, and a single phylogeny may reflect several alignments. For example, alternative alignments might reflect different evolutionary histories of the characters, but they could all produce the same phylogeny when analyzed. There is, thus, an asymmetry between alignments and phylogenies, rather than the symmetrical relationship implied by the usual notion of interchangeability of trees and alignments.

A second important point is that groups of contiguous nucleotides on a chromosome are typically descended together from a common ancestor. That is, nucleotides are rarely singletons with respect to their phylogenetic history, but occur as non-recombining sequence blocks. The sequence blocks can be rearranged owing to recombination and translocation, but they usually function as a unit, for example, as part of a protein-coding gene, an intron, a structural RNA, a transcribed spacer or a regulatory mi-RNA. Conceptually, evolutionary homology applies directly to these loci, even though it may be difficult to apply this concept in practice (i.e. the region boundaries may be indeterminable).

Standard statistical methods are not effective at extracting phylogenetic information from sequence data when each position is subject to a common process, because the IID assumptions are violated. This is the problem with current edit models. Nucleotides are subject to a common process *within* sequence block but to many different processes *between* blocks. Alignment must recognize blocks of nucleotides as characters, as well as single nucleotides.

A third valuable point is that, in order to study homologies, we need to move the focus from the sequences as a contiguous string of nucleotides along a chromosome (the alignment rows) to the evolutionary characters (the alignment columns). That is, current alignment algorithms usually work horizontally along the tabulated alignment, but the alignment itself is interpreted vertically. Remember, character homology is a relationship among the nucleotides columns of the table, not among the rows — the rows are the objects and columns are the characters.

So, the fundamental practical limitation of all current computer algorithms for multiple-sequence alignment is that they focus on the rows not the columns! A multiple alignment is not simply a set of pairwise alignments braided together — it is the “multiple” part of the procedure that is of most interest for homology, not the “pairwise” part. Manual methods of multiple sequence alignment all work vertically, and this is their great strength. It is the principal reason why manual “adjustment” is so frequently employed by phylogeneticists[6] — the homology assessments as produced by the computer program, working along the rows, are re-evaluated by looking at the patterns across the rows within the columns. Misaligned homology is often *very* obvious when the alignment is inspected vertically. This approach has even been formalized.[83]

As a fourth point, I have now made it clear when and why current computerized algorithms do not succeed in consistently detecting homologous nucleotides — each computer program generally relies on a single criterion for inferring homology. For example, optimal similarity alignments fail because they restrict themselves to the criterion of compositional similarity; and direct optimization fails because it restricts itself to the criterion of congruence.

To make these criteria operational, we need to compare their inferences by evaluating the comparative evidence, not focussing on one criterion alone. Most of the relevant ideas have actually been implemented singly in different alignment programs. What has not happened is anyone trying to put them together in some coherent way, so that they can be compared as evidence for homology.

Morrison et al.[11] (p. 56) provides a summary of the most commonly used programs for multiple sequence alignment in phylogenetics, listing which criterion, or occasionally pairs of criteria, they try to implement. We now need a program that implements all of the criteria simultaneously.

Morrison[20] discusses several possible approaches to the computational problem, including the following:

- try to reproduce the human approach to homology assessment, which is by proposing homology hypotheses based on similarity and conjunction, which are then tested with congruence;
- search the nucleotide sequences for evidence of known molecular processes, and then optimize the combination of these to produce a set of optimal scenarios for the origin of the sequence variation;
- evaluate the types of similarity independently as the criteria for alignment hypotheses, represent the hypotheses as a (large) set of local alignments, and then combine these local alignments into a global alignment;
- use as a starting point a pre-existing alignment that has been curated by phylogeneticists (and trusted by them), and then add new sequences to it — this allows the high quality of the initial alignment to be maintained as the alignment grows in size;
- use as a suitable starting point an alignment based on compositional similarity, which will identify highly conserved sequence blocks, and then modify it to represent a scenario of postulated homologies — this is apparently what is currently being undertaken manually by many biologists, using some convenient visualization tool;[84]
- try to improve the current heuristic algorithms, possibly with a biological justification (e.g. variable scoring matrices[85]) but without explicitly addressing homology — this is apparently what is currently being undertaken by many bioinformaticians.

To this list I would add the concepts of pre- and post-processing the sequences. That is, it might be possible to pre-process the sequences to extract information relevant to homology detection. In some ways, this is the approach used when creating the so-called homology matrices for amino-acid sequences (e.g. BLOSUM (Henikoff), PAM (Dayhoff) or GONNET). Post-processing has been popular for multiple sequence alignment, in which an initial alignment created under one criterion is then modified using a different criterion (e.g. using iterative refinement[86]).

As an aside, I will point out that there are only three practical strategies currently used by phylogeneticists for multiple sequence alignment:[6] (1) try to use the available computer programs developed for one general sequence comparison (e.g. CLUSTAL, MAFFT, ProbCons); (2) combine alignment and tree-building, via statistical alignment or direct optimization; or (3) manually intervene in the alignment process, in an attempt to correct obvious “mistakes” with respect to likely homology.

Clearly, if inferences from the different criteria for homology contradict each other then we will have ambiguous alignments. That is, there will be more than one plausible alternative alignment. Any automated procedure for multiple sequence alignment must be able to resolve these conflicts in some biologically acceptable manner. Mathematically, this is usually done by assigning weights (or costs) to the different inferences, so that this cost can be optimized for each data set,[68] or a probability function is used based on a joint probability model.[87]

Finally, nothing that I have said here excludes the possibility that for some, if not many, sequence regions homology will be indeterminable. These are the so-called "unalignable" regions that most practitioners will have encountered at some time,[88] for which homology assessment remains intractable. If any one sequence region cannot be aligned across all of the sequences, then that block can be presented as several consecutive subalignments, with each group of aligned sequences being offset horizontally from the others in a staggered manner.[3] This method preserves all of the available homology information within each subalignment, without falsely aligning non-homologues. This will be a better practice than abandoning the information, as is the case when gapped regions are excluded from the final alignment.[89]

## Conclusion

I once commented that:[90]

> "The basic problem with sequence alignment is that it seems to be more an art than a science. For a science, the techniques are scarcely rigid enough, whereas for an art the results are usually rather prosaic. Perhaps, it is most justly treated as a sport, one for which no universal rules are presently formulated."

I have not changed my opinion since then. So, in this paper I have suggested ways in which the current "rules" could be changed to make the "games" more relevant to phylogeneticists. With a bit of luck, a few bioinformaticians will be inspired to implement some of these new rules.

## Acknowledgements

Thanks to Scot Kelchner for helpful discussions over many years. Thanks also to Försäkringskassan for funding.

## Declaration of Conflicting Interests

The author has declared no potential conflicts of interest with respect to the research, authorship, and/or publication of this article.

Figure legends

**Figure 1.**

Schematic insect body (left), showing body segments, legs, and mouthparts. A 5-step scenario (right) showing the historical (evolutionary) origin of the segments, legs, and mouthparts. The three segments are homologous, as are the legs and mouthparts. The schematic shows a top view of the insect, while the scenario shows side views.

**Figure 2.**

The hierarchical nature of homology, as exemplified by forelimbs.

**Figure 3.**

An example of a molecular hierarchy of homology. Each row represents a level of the hierarchy, with the parts in the left column, and the molecular process that connect the levels in the right column.

**Figure 4.**

Partial alignment of the 70-kDa heat-shock protein (Hsp70) gene for nine species of the phylum Apicomplexa. The nucleotides are colour coded based on their translated amino acids. Reproduced from ref.[11]

**Figure 5.**

Part of a nucleotide sequence alignment of chloroplast DNA from 24 species of bamboos (from ref.[91]). Seven different events in the evolution of the sequence variation are numbered, and the subsequences involved are boxed. Also shown is the stem-loop structure of the transcribed RNA.

Schematic insect body

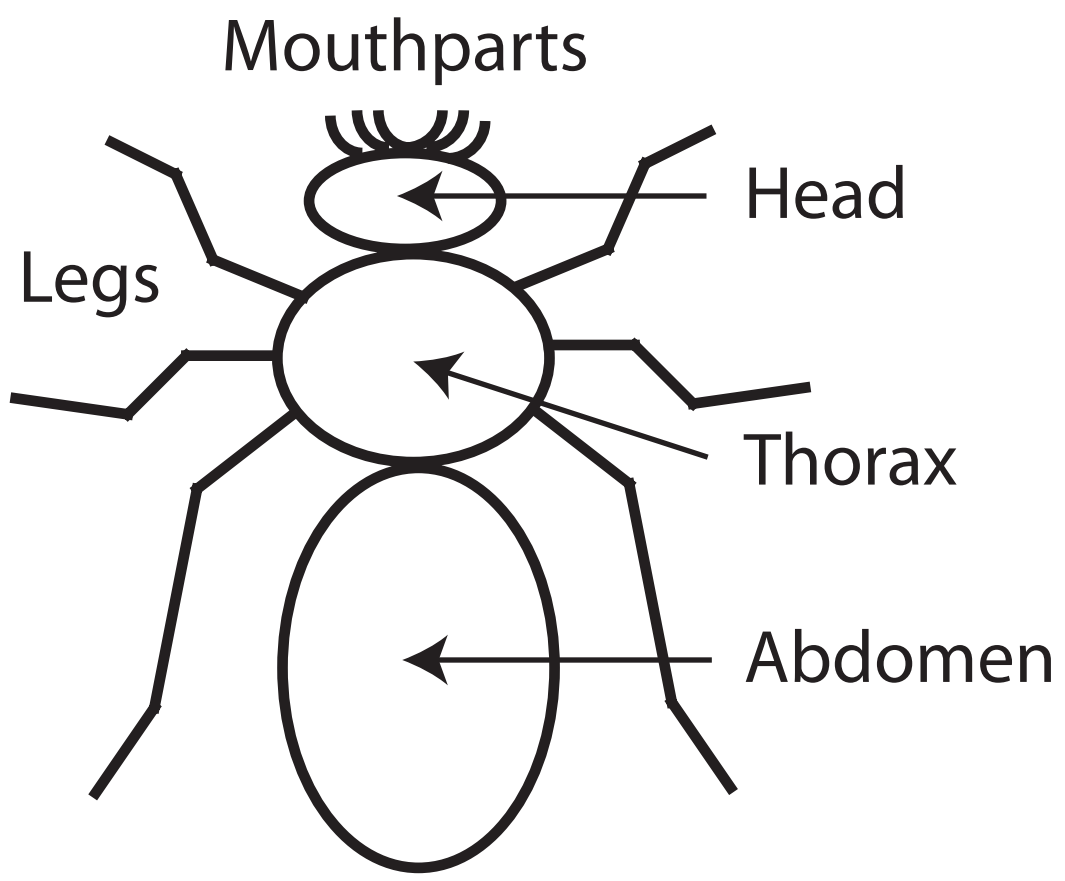

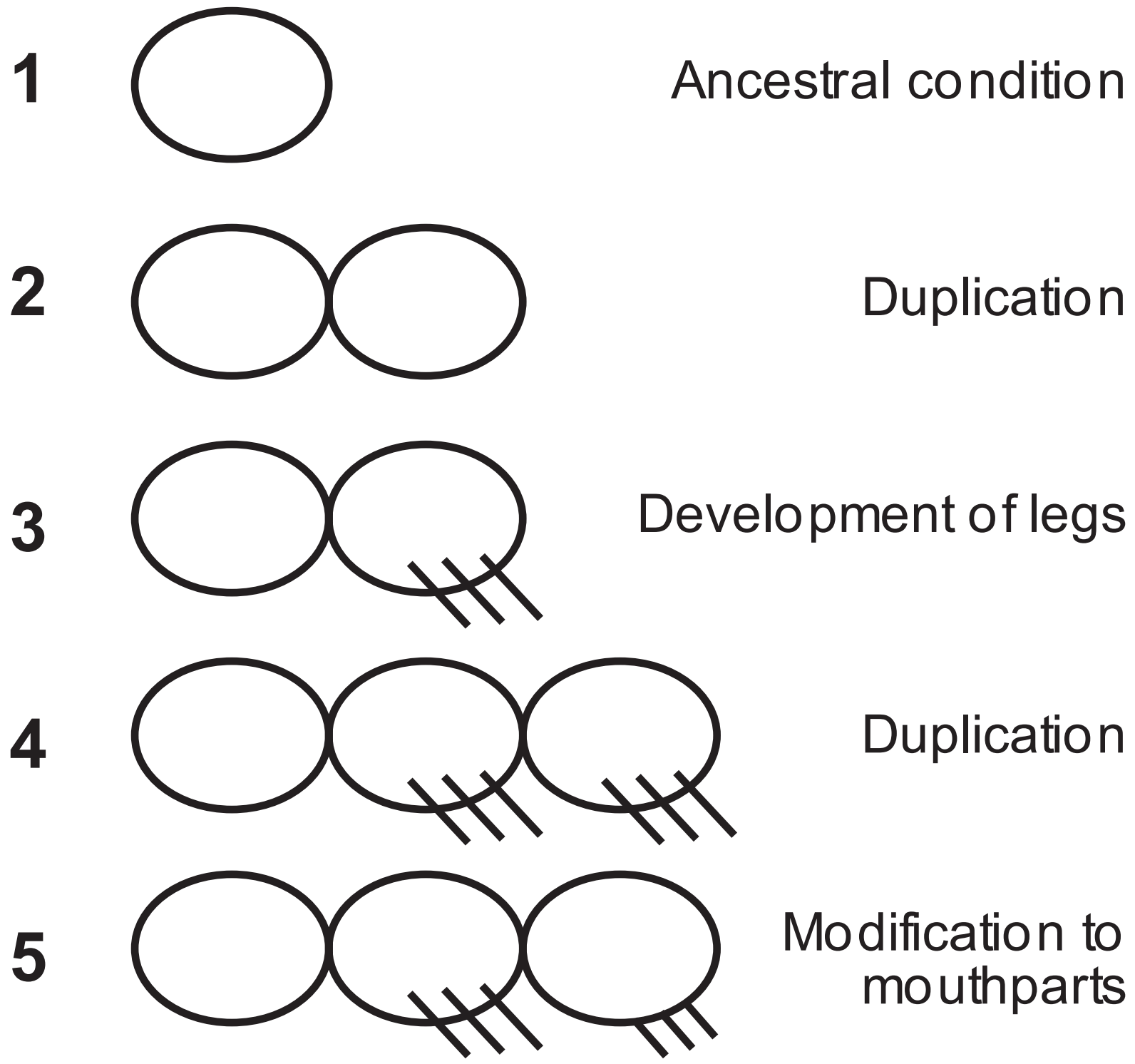

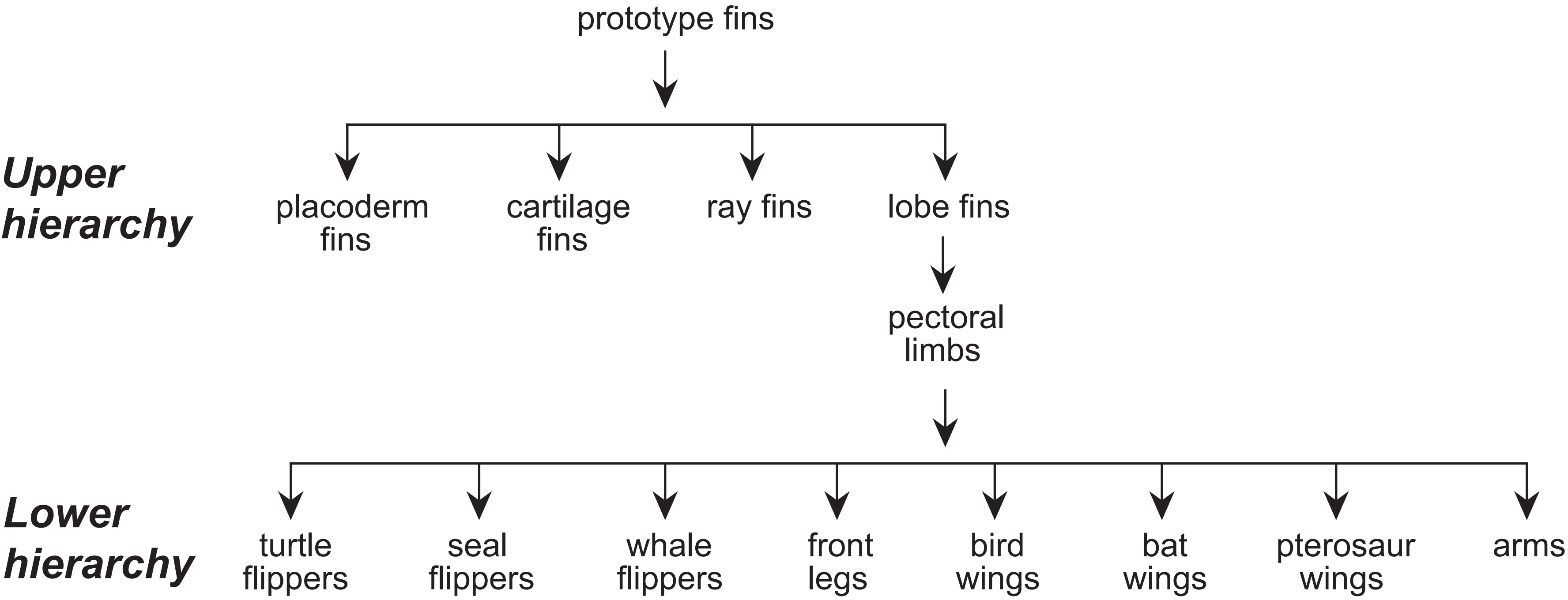

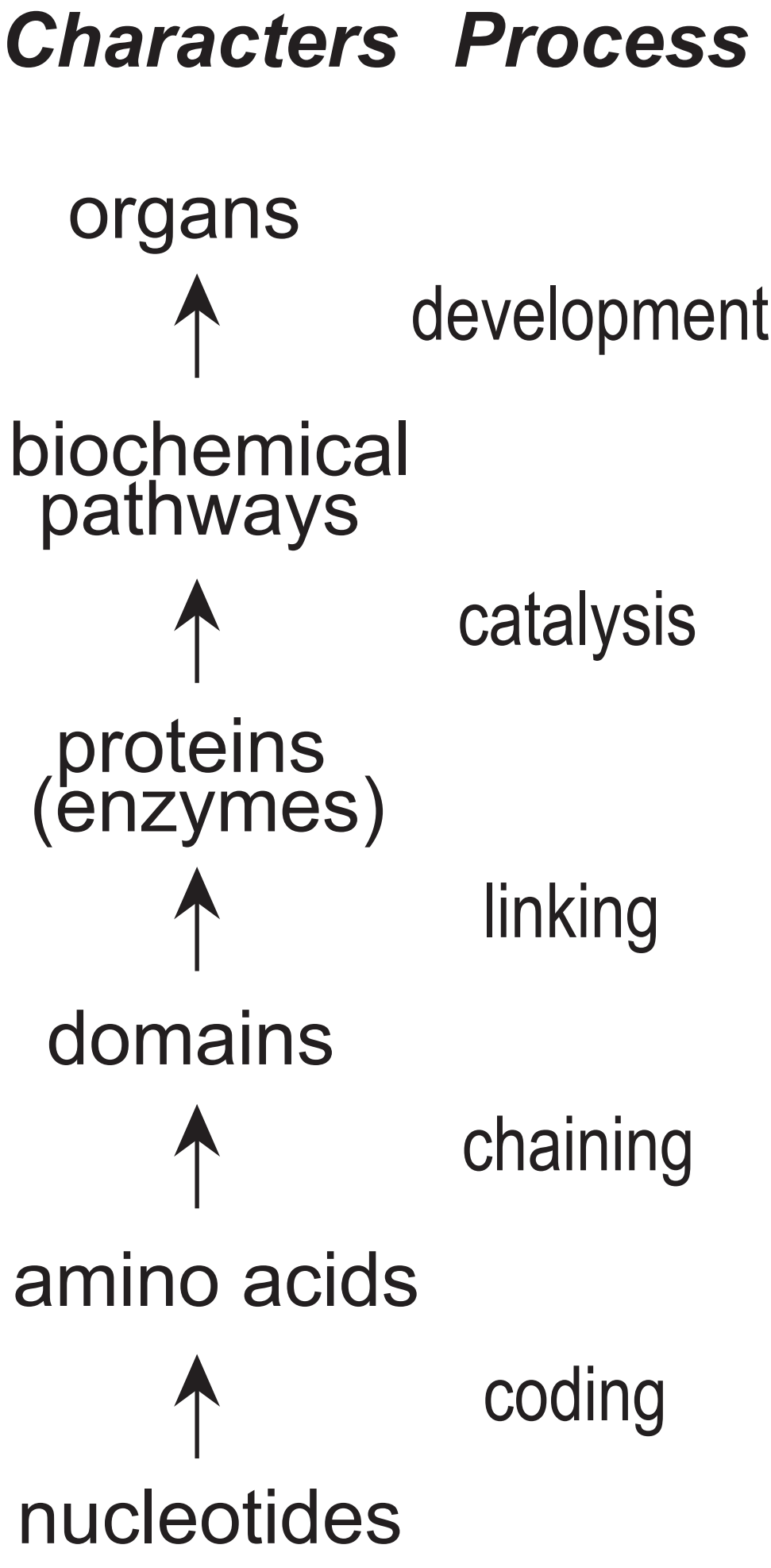

| | |
|---|---|
| *Plasmodium falciparum* | GATGTTTGCTCC–TTATCA |
| *Babesia bovis* | GATGTC–GCTCCACTCTCC |
| *Cryptosporidium parvum* | GATGTT–GCTCCATTATCA |
| *Cryptosporidium canis* | GATGTT–GCCCCACTGTCC |
| *Cryptosporidium baileyi* | GATGTT–GCTCCATTATCA |
| *Cryptosporidium felis* | GATGTT–GCTCCTTTGTCT |
| *Cryptosporidium wrairi* | GATGTT–GCTCCATTATCA |
| *Cryptosporidium saurophilum* | GATGTT–GCTCCTTTGTCT |
| *Cryptosporidium meleagridis* | GATGTT–GCTCCATTATCA |

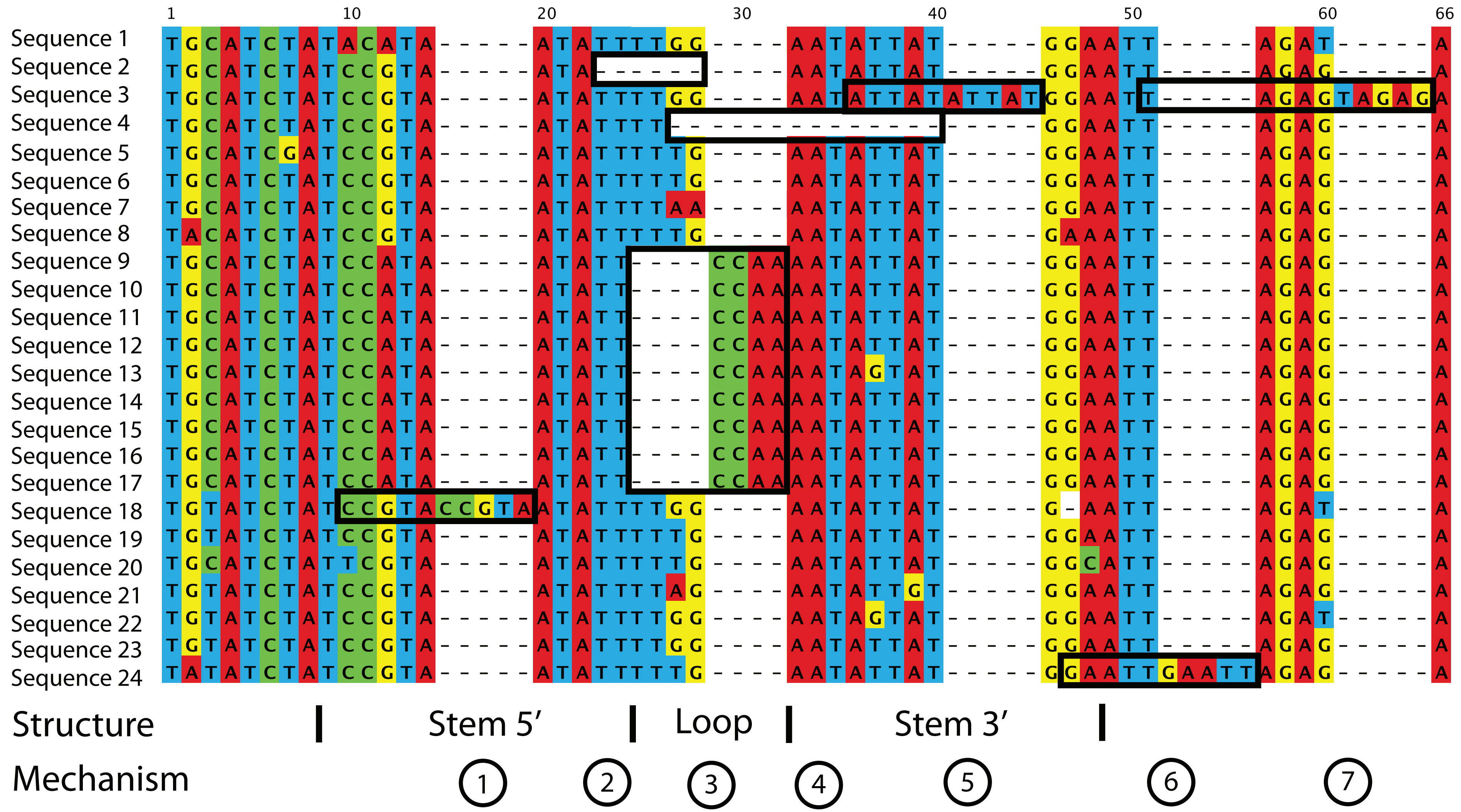